\documentclass[%
preprint,
amsmath,
amssymb,
aps,
nofootinbib,
12pt,
prd
]{revtex4-1}

\usepackage{graphicx}
\usepackage{dcolumn}
\usepackage{bm}
\usepackage{hyperref}

\hypersetup{pdftitle={Kinematic Numerators and a Double Copy Formula for N=4 Super-Yang‒Mills Residues},
 pdfauthor={Sean Litsey and James Stankowicz},
 pdfsubject={High Energy Theoretical Physics},
 pdfkeywords={Scattering Amplitudes}}

\begin{document}

\title{Kinematic Numerators and a Double-Copy Formula for \texorpdfstring{$\mathcal{N}=4$}{N=4}
Super-Yang-Mills Residues}

\author{Sean Litsey and James Stankowicz}

\date{September 29, 2013}

\affiliation{Department of Physics and Astronomy,\\
University of California, Los Angeles \\
Los Angeles, CA 90095-1547, USA}
\begin{abstract}
Recent work by Cachazo, He, and Yuan shows that connected prescription
residues obey the global identities of $\mathcal{N}=4$ super-Yang-Mills
amplitudes. In particular, they obey the Bern-Carrasco-Johansson (BCJ)
amplitude identities. Here we offer a new way of interpreting this
result via objects that we call \emph{residue numerators.} These objects
behave like the kinematic numerators introduced by BCJ except that
they are associated with individual residues. In particular, these
new objects satisfy a double-copy formula relating them to the residues
appearing in recently-discovered analogs of the connected prescription
integrals for $\mathcal{N}=8$ supergravity. Along the way, we show
that the BCJ amplitude identities are equivalent to the consistency
condition that allows kinematic numerators to be expressed as amplitudes
using a generalized inverse.
\end{abstract}
\maketitle

\section{Introduction}

There has been much recent progress in calculating scattering amplitudes
in gauge and gravity theories. Among the many advances is the discovery
by Bern, Carrasco, and Johansson (BCJ) of a duality between color
and kinematics~\cite{bern_new_2008}. In this duality, kinematic
numerators of diagrams obey relations similar to the Jacobi identities
obeyed by color factors. One consequence of these relations is that
color-ordered tree-level partial amplitudes obey a set of nontrivial
identities, known as BCJ amplitude identities. These amplitude identities
are even more constraining than the Kleiss-Kuijf identities~\cite{kleiss_multigluon_1989},
and so reduce the number of partial amplitudes required to determine
a full scattering amplitude. Color-kinematic duality has been conjectured
to hold for any number of loops or legs in a wide variety of theories,
including pure Yang-Mills theories and their supersymmetric extensions.
While no proof exists at loop level, a variety of nontrivial constructions
exist~\cite{bern_color-kinematics_2013,bern_perturbative_2010,bern_simplifying_2012,carrasco_one-loop_2013,boels_colour-kinematics_2013,carrasco_five-point_2012,bjerrum-bohr_integrand_2013,boels_color-kinematic_2013},
and its original tree-level formulation has been fully proven.

One remarkable feature of color-kinematic duality is that one can
use it to construct gravity amplitudes directly from gauge-theory
amplitudes. To do so one replaces the color factors in a gauge-theory
amplitude with corresponding kinematic numerators that obey the duality.
This gives what is known as the double-copy form of gravity amplitudes~\cite{bern_perturbative_2010,bern_new_2008}.
At tree level, this encodes the Kawai-Lewellen-Tye (KLT) relations~\cite{kawai_relation_1986}
between gauge and gravity amplitudes~\cite{bern_gravity_2010}. The
extension of the double-copy formula to loop level~\cite{bern_perturbative_2010}
requires first constructing loop-level Yang-Mills amplitudes in a
form where the duality is manifest. Then, as at tree level, one obtains
gravity amplitudes simply by replacing color factors with corresponding
kinematic-numerator.

A parallel direction of research originated from writing the tree-level
scattering amplitude of $\mathcal{N}=4$ super-Yang-Mills with the
Roiban, Spradlin, Volovich, and Witten (RSVW) twistor string formula~\cite{witten_perturbative_2004,berkovits_alternative_2004,roiban_googly_2004,roiban_tree-level_2004,spradlin_twistor_2009}.
(This is also known as the ``connected prescription''.) The RSVW
formula expresses the scattering amplitude as an integral over a moduli
space of curves in $\mathbb{CP}^{3|4}$ supertwistor space, effectively
reducing the entire scattering amplitude calculation to solving an
algebraic system of equations.

This method for determining tree-level scattering amplitudes as integrals
was extended to $\mathcal{N}=8$ supergravity, first in specific cases~\cite{nguyen_tree_2010,hodges_new_2011,hodges_simple_2012},
and later in general~\cite{cachazo_twistor_2012,cachazo_gravity_2013,cachazo_gravity_2012,he_link_2012}.
Like the RSVW formula, the integrals for $\mathcal{N}=8$ supergravity
can be interpreted as contour integrals, and hence as sums of residues.
The formula appearing in Ref\@.~\cite{cachazo_twistor_2012} was
originally a conjecture, in part because the formula required the
KLT relations to hold for the residues in the same way that it holds
for the amplitudes; this ``KLT orthogonality conjecture'' has now
been proven~\cite{cachazo_scattering_2013-2}.

Very recently, $\mathcal{N}=4$ super-Yang-Mills amplitudes have been
constructed using the newfound ``scattering equations''~\cite{cachazo_scattering_2013},
shown to have a color-kinematic structure, and have been explicitly
related to color-dual numerators~\cite{cachazo_scattering_2013-3}.
Because both the scattering-equation-based amplitudes and the kinematic
numerator decomposition of amplitudes yield identical scattering amplitudes,
it is not surprising that the former can be written in terms of the
latter. On the other hand, the discovery that global relations, such
as the KLT relations, hold at the residue level is intriguing because
it suggests there is a Jacobi-like numerator structure for the residues
as well.

Additional hints of a Jacobi-like numerator structure at the level
of residues appeared in a proof of the BCJ amplitude identities in
$\mathcal{N}=4$ super-Yang-Mills~\cite{cachazo_fundamental_2012}.
The proof uses the explicit structure of the RSVW integrand to prove
the BCJ relations. Further, Ref.~\cite{cachazo_fundamental_2012}
discusses how the method of the proof indicates that the BCJ amplitude
identities hold at the level of the residues themselves, analogously
to the KLT orthogonality conjecture. Given that the original BCJ amplitude
relations were originally derived starting from considering color-dual
numerators, the reappearance of amplitude relations for residues strongly
hints at an analogous set of numerators for the residues.

Following these hints, we define objects called \emph{residue numerators}.
By construction, the residue numerators are analogous to the kinematic
numerators of partial amplitudes, except that they hold for RSVW residues.
We use the KLT orthogonality conjecture to show that the residue numerators
obey both a double-copy formula and an orthogonality condition. We
also expound on the observation of Ref\@.~\cite{cachazo_fundamental_2012}
to verify that RSVW residues do indeed satisfy the BCJ relations.
To do all this, we work in the linear algebra formalism of Refs\@.~\cite{vaman_constraints_2010,boels_powercounting_2013}.
To formally prove our results in this formalism, we prove a conjecture
of Ref.~\cite{boels_powercounting_2013}, establishing that the BCJ
amplitude identities are equivalent to a consistency condition equation.

The structure of the paper is as follows. Section~\ref{sec:Background-Material-and-Lemmas}
outlines the necessary background material: the linear algebra formalism,
the RSVW formula and its gravitational generalizations, and the concepts
of color-kinematic duality and double copy. Section~\ref{sec:Background-Material-and-Lemmas}
also contains some novel material. Subsection~\ref{sub:Algebraic-Equivalence}
presents a proof that the BCJ amplitude identities are equivalent
to a system of constraint equations in the linear algebra formalism.
In subsection~\ref{sub:Proof-of-BCJ-for-Residues} we explicitly
demonstrate that the BCJ amplitude identities (and hence the constraint
equations) apply to RSVW residues, as was noted in~\cite{cachazo_fundamental_2012}.
Section~\ref{sec:Residue-Numerators} defines residue numerators,
the central objects of this paper, and proves their double-copy formula.
Finally, Section~\ref{sec:Conclusion-and-Future-Work} provides a
conclusion and discusses future directions.

\section{Background Material and Lemmas\label{sec:Background-Material-and-Lemmas}}

\subsection{Linear Algebra}

A convenient way of structuring discussions of kinematic numerators
is the linear algebra approach pioneered in Ref.~\cite{vaman_constraints_2010}
and extended in Ref.~\cite{boels_powercounting_2013} (whose notation
we largely adopt). This formalism makes generalized gauge invariance
manifest, and, as we shall demonstrate, also reinterprets the BCJ
amplitude identities as algebraic consistency conditions.

To motivate this approach, recall that Yang-Mills scattering amplitudes
in four dimensions can be written in a so-called Del Duca-Dixon-Maltoni
(DDM) decomposition~\cite{del_duca_new_2000}. In this form, the
full $n$-particle amplitude at tree level is written as 
\begin{equation}
\mathcal{A}_{n}=g^{n-2}\sum_{\tau\in S_{n-2}}c_{\tau}A_{n}\left(1,\tau\left(2\right),\dots,\tau\left(n-1\right),n\right),
\end{equation}
where the coupling constant is $g$, the notation $\tau\in S_{n-2}$
indicates that the sum runs over permutations $\tau$ of the particle
labels $2,\,\dots\,,\, n-1$, the $A_{n}$ are color-ordered partial
amplitudes, and the $c_{\tau}$ are color factors%
\footnote{These are a product of group-theory structure constants; see \cite{del_duca_new_2000}
for details. %
} of cubic diagrams. Cubic diagrams are diagrams with only trivalent
vertices, which conserve color and momentum at each vertex. Any diagrams
containing higher-point contact terms are absorbed into cubic diagrams
with the same color factor, with missing propagators $P$ introduced
by multiplying by $1=\frac{P}{P}$. While there is no known Lagrangian
from which this decomposition can be directly generated by Feynman
rules, these trivalent diagrams are a useful way of reorganizing the
usual sum over Feynman diagrams.. 

Another decomposition of the tree level amplitude, which we will refer
to as the BCJ decomposition, is 
\begin{equation}
\mathcal{A}_{n}=g^{n-2}\sum_{i=1}^{\left(2n-5\right)!!}\frac{c_{i}n_{i}}{D_{i}},\label{eq:BCJ decomposition}
\end{equation}
where the sum is now over the unique set of $\left(2n-5\right)!!$
cubic diagrams, with color factors $c_{i}$, products of propagators
$D_{i}$, and so-called ``kinematic numerators'' $n_{i}$. These
last objects are functions only of the external momenta and helicities,
and are not uniquely defined. This is because a generalized gauge
transformation $n_{i}\mapsto n_{i}+\Delta_{i}$, for functions $\Delta_{i}$
that obey 
\begin{equation}
\sum_{i=1}^{\left(2n-5\right)!!}\frac{c_{i}\Delta_{i}}{D_{i}}=0,\label{eq:Generalized gauge transformation}
\end{equation}
will leave the BCJ decomposition Eq.~(\ref{eq:BCJ decomposition})
invariant. The notion of a generalized gauge transformation will turn
out to be nicely expressed in the linear algebra formalism.

We can relate the DDM and BCJ decompositions in a useful way. It was
shown in Ref\@.~\cite{del_duca_new_2000} that the $\left(n-2\right)!$
color factors $c_{\tau}$ form a basis of the space of color factors
of cubic diagrams. This is possible because the Jacobi relations of
the structure constants induce linear relations among the color factors.
In other words, any of the $\left(2n-5\right)!!$ color factors $c_{i}$
that appear in the decomposition Eq.~(\ref{eq:BCJ decomposition})
can be written as
\begin{equation}
c_{i}=\sum_{\tau\in S_{n-2}}W_{i\tau}c_{\tau}\label{eq:color basis}
\end{equation}
where $W_{i\tau}$ is a $\left(2n-5\right)!!\times\left(n-2\right)!$
matrix that encodes the Jacobi relations among the color factors.
Our notation expresses sums over permutations (as in the DDM decomposition)
with $\tau$ and $\omega$, and sums over cubic diagrams by Latin
indices $i$, $j$.

Color-kinematic duality states that there exists a set of color-dual
numerators $n_{i}$ that obey the exact same Jacobi relations as the
color factors $c_{i}$. In other words, for the same matrix $W_{i\tau}$
defined above in Eq.~(\ref{eq:color basis}), we can write 
\begin{equation}
n_{i}=\sum_{\tau\in S_{n-2}}W_{i\tau}n_{\tau}\label{eq:numerator basis}
\end{equation}
for some set of $\left(n-2\right)!$ numerators $n_{\tau}$. Substituting
Eq.~(\ref{eq:color basis}) and Eq.~(\ref{eq:numerator basis})
into the BCJ decomposition, we find
\begin{eqnarray}
\mathcal{A}_{n} & = & g^{n-2}\sum_{i=1}^{\left(2n-5\right)!!}\sum_{\tau,\omega\in S_{n-2}}\frac{c_{\tau}n_{\omega}}{D_{i}}W_{i\tau}W_{i\omega}\nonumber \\
 & = & g^{n-2}\sum_{\tau,\omega\in S_{n-2}}c_{\tau}n_{\omega}F_{\tau\omega},\label{eq: full amplitude in terms of F}
\end{eqnarray}
where $F_{\tau\omega}$ is an $\left(n-2\right)!\times\left(n-2\right)!$
symmetric matrix with products of inverse propagators as entries:
\begin{equation}
F_{\tau\omega}\equiv\sum_{i=1}^{\left(2n-5\right)!!}\frac{W_{i\tau}W_{i\omega}}{D_{i}}.
\end{equation}
The matrix $F_{\tau\omega}$ is a convenient way of simultaneously
encoding both the color and numerator Jacobi relations in the basis
of partial amplitudes.

Equating Eq.~(\ref{eq: full amplitude in terms of F}) to the DDM
decomposition and matching coefficients of the $c_{\tau}$, we have
the identity
\begin{equation}
A_{n}\left(1,\tau\left(2\right),\dots,\tau\left(n-1\right),n\right)=\sum_{\omega\in S_{n-2}}F_{\tau\omega}n_{\omega}.
\end{equation}
This can be thought of in matrix notation as a system of linear equations
\begin{equation}
FN=A\label{eq:Fn=00003DA}
\end{equation}
in the $\left(n-2\right)!$-dimensional space of partial amplitudes
spanned by Kleiss-Kuijf basis amplitudes and indexed by $\tau\in S_{n-2}$,
and $N$ is a column vector of numerators. Ideally we could invert
this formula to get an expression for the numerators in terms of the
partial amplitudes, but this is impossible because $F$ is singular.
This is no surprise: the invariance of the full amplitude under generalized
gauge transformations Eq.~(\ref{eq:Generalized gauge transformation})
ensures that the numerators are not unique, so $F$ cannot be invertible.

Therefore $F$ has a nontrivial kernel. To circumvent this problem,
Ref.~\cite{boels_powercounting_2013} suggested using the machinery
of generalized inverses (also called pseudoinverses). A generalized
inverse is a matrix $F^{+}$ satisfying $FF^{+}F=F$, and it can be
shown that such an $F^{+}$ always exists, but is not unique. Generalized
inverses are useful because of the following theorem~\cite{ben-israel_generalized_2003}:
a solution to $FN=A$ exists if and only if the consistency condition
\begin{equation}
FF^{+}A=A\label{eq:consistency condition}
\end{equation}
holds for some generalized inverse $F^{+}$. The general solution
is then given by
\begin{equation}
N=F^{+}A+\left(I-F^{+}F\right)v
\end{equation}
for an arbitrary vector $v$ that parametrizes the kernel of $F$.

Notice that $I-F^{+}F$ is a projection operator onto the kernel of
$F$, since, by the definition of $F^{+}$, $F\left(I-F^{+}F\right)=F-FF^{+}F=0$.
The consistency condition Eq.~(\ref{eq:consistency condition}) has
been conjectured to be equivalent to the BCJ amplitude identities~\cite{boels_powercounting_2013},
and we prove this conjecture below. Note, however, that because the
existence of color-dual numerators has been proven at tree level~\cite{mafra_explicit_2011,bjerrum-bohr_momentum_2011},
we know that the consistency condition is satisfied thanks to the
``if and only if'' logic. Our proof that the consistency condition
Eq.~(\ref{eq:consistency condition}) is equivalent to the BCJ amplitude
identities gives an alternative proof of the existence of color dual
numerators, one that will extend to the residue numerators defined
in Section 3.

\subsubsection{Equivalence of the BCJ Amplitude Identities and the Consistency Condition\label{sub:Algebraic-Equivalence}}

Our goal is to show that $FF^{+}A=A$ is the same as the BCJ amplitude
identities, which we shall write formally as $\mathcal{S}A=0$, where
$\mathcal{S}$ is a matrix that forms the linear combination of amplitudes
appearing in the BCJ identities. Since we chose the $A$ in $FF^{+}A$
to be a vector of amplitudes in the Kleiss-Kuijf basis, we have to
be careful to choose the matrix $\mathcal{S}$ so that it acts on
the same vector of Kleiss-Kuijf amplitudes to produce the BCJ amplitude
identities. One particularly nice choice for $\mathcal{S}$ that accomplishes
this is a matrix representation of the momentum kernel~\cite{bjerrum-bohr_gravity_2010,bjerrum-bohr_proof_2010,bjerrum-bohr_new_2010,bjerrum-bohr_momentum_2011},
given explicitly by $\mathcal{S}_{\tau\omega}=\mathcal{S}\left[\tau\left(2\right),\tau\left(3\right),\dots,\tau\left(n-1\right)|\omega\left(2\right),\omega\left(3\right),\dots,\omega\left(n-1\right)\right]$.
This matrix has polynomials of Mandelstam variables as entries, and
many of its properties are discussed in Ref.~\cite{bjerrum-bohr_momentum_2011}.

There are two assumptions that will go into our proof. These assumptions
are both widely believed and have survived extensive low-point checks,
but remain unproven for general $n$. The first is that the basis
of $\left(n-3\right)!$ amplitudes appears to be minimal~\cite{bjerrum-bohr_minimal_2009,bjerrum-bohr_algebras_2012,stieberger_open_2009}.
In other words, there are not further relations that will reduce the
number of independent amplitudes below the $\left(n-3\right)!$ BCJ-independent
amplitudes. The second assumption is that Eq.~(\ref{eq:Fn=00003DA})
only has solutions for $A$ in the basis spanned by the $\left(n-3\right)!$
BCJ-independent amplitudes. It is known that the span of these amplitudes
is sufficient for a wide class of theories, including both Yang-Mills
theories and the colored trivalent scalar theories discussed in \cite{bjerrum-bohr_algebras_2012}.
Ourassumption is that this sufficiency holds for any theory that possesses
color-kinematic duality, and so this sufficiency also holds for any
theories for which $F$ can be defined.

We demonstrate that the two equations $FF^{+}A=A$ and $\mathcal{S}A=0$
impose the same constraints on the elements of $A$ using a simple
dimension counting argument. As mentioned above, the BCJ amplitude
identities are known to reduce the $\left(n-2\right)!$ Kleiss-Kuijf
independent amplitudes to $\left(n-3\right)!$ independent amplitudes~\cite{bern_new_2008,bjerrum-bohr_minimal_2009,stieberger_open_2009},
i.e. the rank of $\mathcal{S}$ is $\left(n-2\right)!-\left(n-3\right)!$.
Our two assumptions imply that for generic momenta, the solution space
of Eq.~(\ref{eq:Fn=00003DA}) necessarily has dimension $\left(n-3\right)!$
i.e.~$\mbox{rank}\, F=\left(n-3\right)!$. But $\mbox{rank}\left(FF^{+}\right)=\mbox{rank}\left(F\right)$
by a theorem of linear algebra~\cite{ben-israel_generalized_2003},
so the image of $FF^{+}$ has dimension $\left(n-3\right)!$. Then
the solution space of $FF^{+}A=A$ has dimension at most $\left(n-3\right)!$,
since it must be contained in the image of $FF^{+}$. But then by
the assumption that the $\left(n-3\right)!$ basis is minimal, the
solution space of $FF^{+}A=A$ must have dimension equal to $\left(n-3\right)!$.
The basis vectors of this space must therefore be BCJ basis amplitudes,
up to at most a linear transformation. Therefore the operators $FF^{+}$
and $\mathcal{S}+I$ must be equal up to this linear transformation,
proving the result. Explicitly, we write $Q\left(FF^{+}-I\right)=\mathcal{S}$
for some $Q\in GL\left(\left(n-3\right)!\right)$ embedded in an $\left(n-2\right)!\times\left(n-2\right)!$
matrix with all other entries zero, and $I$ the $\left(n-2\right)!\times\left(n-2\right)!$
identity matrix.

This may be understood geometrically. If the BCJ amplitudes form the
true minimal basis of color-ordered amplitudes, then the larger space
spanned by Kleiss-Kuijf amplitudes must be constrained to the smaller
space spanned by the BCJ amplitudes. We illustrate this for the four-point
case in Fig.~\ref{fig:KK to BCJ}. In this simplest example, there
are $\left(4-2\right)!=2$ amplitudes in the Kleiss-Kuijf basis, and
$\left(4-3\right)!=1$ amplitude in the BCJ basis. If the Kleiss-Kuijf
basis were minimal, then the vector 
\begin{equation}
A=\left(A\left(1,2,3,4\right),A\left(1,3,2,4\right)\right)
\end{equation}
in the plane%
\footnote{We say ``plane'' for $\mathbb{C}^{2}$ and ``line'' for $\mathbb{C}$
to highlight the geometry.%
} $\mathbb{C}^{2}$ would fully determine all partial amplitudes. The
four-point BCJ basis linearly relates the two elements of the vector
$A$ by 
\begin{equation}
A\left(1,3,2,4\right)=\frac{u}{s}A\left(1,2,3,4\right)\textrm{ or }A\left(1,2,3,4\right)=\frac{s}{u}A\left(1,3,2,4\right)
\end{equation}
where either equation is valid, and amounts to choosing either $A\left(1,2,3,4\right)$
or $A\left(1,3,2,4\right)$ as a basis amplitude. This is equivalent
to projecting to one axis or the other in the Fig.~\ref{fig:KK to BCJ}.
Since the Kleiss-Kuijf vectors $A$ describe physically valid partial
amplitudes, they cannot lie at an arbitrary point in the plane, but
must instead lie on the \emph{BCJ line}. For the four-point case,
both operators $FF^{+}$ and $\mathcal{S}+I$ are rank one, and act
on the Kleiss-Kuijf vector of amplitudes. This means both operators
necessarily map to a point along the BCJ line. The linear transformation
$Q\in GL\left(\left(4-3\right)!\right)$ in this case is just a constant
that translates a point along the BCJ line, but such movement does
not alter the relation between the amplitudes. The next-highest-point
case, $n=5$, has $\left(5-2\right)!=6$ amplitudes in the Kleiss-Kuijf
basis and $\left(5-3\right)!=2$ amplitudes in the BCJ basis; geometrically
the $n=5$ case corresponds to a $\mathbb{C}^{6}$ hyperplane for
the Kleiss-Kuijf basis with all points actually lying on the $\mathbb{C}^{2}$
plane spanned by the BCJ basis amplitudes. This same line of geometric
reasoning supports the rank-counting argument for all $n$.

\begin{center}
\begin{figure}
\includegraphics{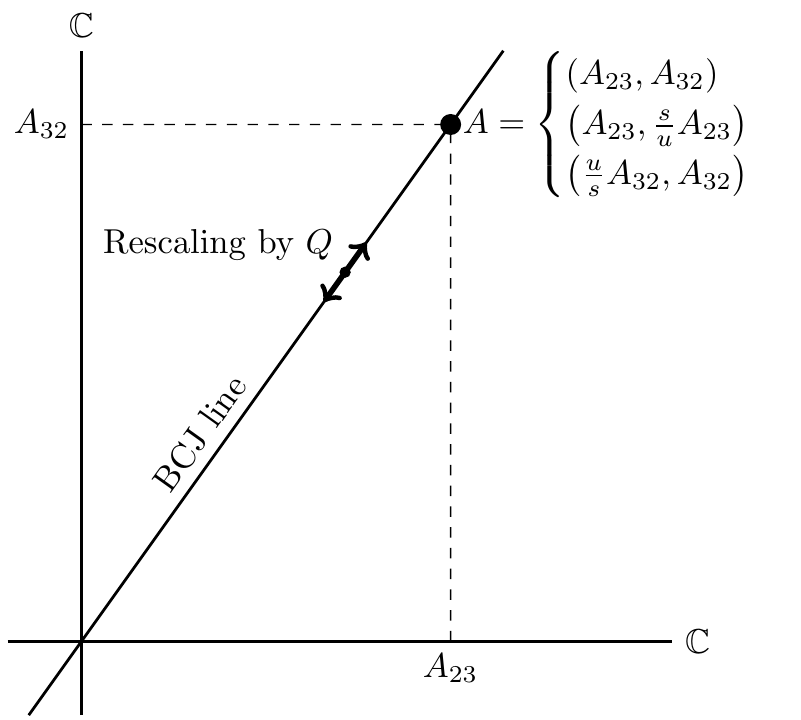}

\caption[Reduction of the Kleiss-Kuijf amplitude basis to the BCJ amplitude
basis for $n=4$.]{\label{fig:KK to BCJ}Reduction of the Kleiss-Kuijf amplitude basis
to the BCJ amplitude basis for $n=4$. In the figure, $A\left(1,2,3,4\right)\equiv A_{23}$
and $A\left(1,3,2,4\right)\equiv A_{32}$. Because the BCJ basis is
the minimal basis, any Kleiss-Kuijf amplitude vector actually lies
on the ``BCJ line''. Both $A_{23}$ and $A_{32}$ are complex numbers,
indicated by the $\mathbb{C}$ labels on the axes. The ``rescaling
by $Q$'' arrows indicate the $GL\left(1\right)$ freedom that rescales
the point $A$ along the BCJ line. }
\end{figure}

\par\end{center}

\subsection{RSVW Formula and Residues}

As mentioned in the introduction, significant work has been done on
the special case of $\mathcal{N}=4$ super-Yang-Mills. In particular,
the aforementioned RSVW formula that gives all tree level partial
amplitudes is 
\begin{eqnarray}
A_{n}\left(1,2,\dots,n\right) & = & \int\frac{d^{2n}\sigma}{\mathrm{vol}\, GL\left(2\right)}\frac{1}{\left(12\right)\left(23\right)\cdots\left(n1\right)}\prod_{\alpha=1}^{k}\delta^{2}\left(C_{\alpha a}\tilde{\lambda}_{a}\right)\delta^{0|4}\left(C_{\alpha a}\tilde{\eta}_{a}\right)\nonumber \\
 &  & \times\int d^{2}\rho_{\alpha}\prod_{b=1}^{n}\delta^{2}\left(\rho_{\beta}C_{\beta b}-\lambda_{b}\right),\label{eq:RSVW no permutation}
\end{eqnarray}
where the $C_{\alpha a}$ are $k\times n$ matrices parametrized by
$\sigma$, as discussed in the Grassmannian formulations of Refs.~\cite{arkani-hamed_duality_2010,arkani-hamed_unification_2011,arkani-hamed_scattering_2012},
for particles in the $R$-charge sector given by $k$. The minors
$\left(12\right)$, $\left(23\right)$, etc. are minors of $C_{\alpha a}$,
and are thus functions of the $\sigma$. The delta functions enforce
the conditions that the spinor helicity variables $\lambda$ and $\tilde{\lambda}$
(along with $\tilde{\eta}$) are appropriately orthogonal, and thus
that overall supermomentum is conserved.

Notice that both the delta functions and the measure are invariant
under permutations of the particle labels. This means we can write
\begin{eqnarray}
A_{n}\left(1,\tau\left(2\right),\dots,\tau\left(n-1\right),n\right) & = & \int\frac{d^{2n}\sigma}{\mathrm{vol}\, GL\left(2\right)}L_{\tau}\prod_{\alpha=1}^{k}\delta^{2}\left(C_{\alpha a}\tilde{\lambda}_{a}\right)\delta^{0|4}\left(C_{\alpha a}\tilde{\eta}_{a}\right)\nonumber \\
 &  & \times\int d^{2}\rho_{\alpha}\prod_{b=1}^{n}\delta^{2}\left(\rho_{\beta}C_{\beta b}-\lambda_{b}\right),\label{eq:RSVW}
\end{eqnarray}
where 
\begin{equation}
L_{\tau}\equiv\frac{1}{\left(1\,\tau\left(2\right)\right)\cdots\left(\tau\left(n-1\right)\, n\right)\left(n\,1\right)}.\label{eq:inverse minor}
\end{equation}
This representation has an important consequence. To understand the
action of $F$ on $A_{\tau},$ we only need to understand how the
inverse minor factor $L_{\tau}$ will be affected, since $L_{\tau}$
is the only factor in $A_{\tau}$ that depends on the particle label
ordering. We are in particular interested in how the consistency condition
$FF^{+}A=A$ manifests itself in this setting. By the permutation
invariance discussed above, $FF^{+}A=A$ holds provided (in matrix
notation) 
\begin{equation}
FF^{+}L=L\label{eq:residue consistency condition}
\end{equation}
on the support of the delta functions. One goal in the remainder of
this section will be to establish this fact.

The proof employs two lemmas. The first lemma is the claim that $FF^{+}A=A$
and $\mathcal{S}A=0$ are algebraically equivalent, which was shown
in Subsection~\ref{sub:Proof-of-BCJ-for-Residues}. The second lemma
states that the BCJ amplitude identities descend to the level of residues.
Mathematically, this is the statement that $\mathcal{S}A=0$ implies
$\mathcal{S}R_{r}=0$ for all RSVW residues $R_{r}$ (which will be
defined below). This was originally noted in Ref.~\cite{cachazo_fundamental_2012}.
Because Ref.~\cite{cachazo_fundamental_2012} proves that $\mathcal{S}L=0$
(in our notation) on the support of the delta functions, the residue
consistency condition Eq.~(\ref{eq:residue consistency condition})
will follow from these two lemmas. Explicitly, we can rewrite $\mathcal{S}L=0$
as
\begin{equation}
Q\left(FF^{+}-I\right)L=0.
\end{equation}
Multiplying by $Q^{-1}$ and rearranging gives Eq.~(\ref{eq:residue consistency condition}).

\subsubsection{Proof of the BCJ Amplitude Identities for RSVW Residues\label{sub:Proof-of-BCJ-for-Residues}}

Our argument proceeds by invoking several equivalent forms of Eq.~(\ref{eq:RSVW})
found in Section 3.1 of Ref.~\cite{arkani-hamed_unification_2011}.
We begin by noting that in supertwistor space, Eq.~(\ref{eq:RSVW})
is 
\begin{equation}
A_{n}\left(1,\tau\left(2\right),\dots,\tau\left(n-1\right),n\right)=\int\frac{d^{2n}\sigma}{\mathrm{vol}\, GL\left(2\right)}L_{\tau}\prod_{\alpha=1}^{k}\delta^{4|4}\left(C_{\alpha a}\mathcal{W}^{\alpha a}\right),
\end{equation}
with $L_{\tau}$ still defined as in Eq.~(\ref{eq:inverse minor}).
Here the $C_{\alpha a}$ are functions of $\sigma$ as given by the
Veronese map discussed in Ref.~\cite{arkani-hamed_unification_2011}.
This can be cleverly rewritten as 
\begin{equation}
A_{n}\left(1,\tau\left(2\right),\dots,\tau\left(n-1\right),n\right)=\int\frac{d^{k\times n}C_{\alpha a}}{\mathrm{vol}\, GL\left(k\right)}G_{\tau}\left(C\right)\prod_{\alpha=1}^{k}\delta^{4|4}\left(C_{\alpha a}\mathcal{W}^{\alpha a}\right)
\end{equation}
where 
\begin{equation}
G_{\tau}\left(C\right)=\int\frac{d^{2n}\sigma\, d^{k\times k}M}{\mathrm{vol}\, GL\left(2\right)}L_{\tau}\prod_{\alpha=1}^{k}\prod_{a=1}^{n}\delta\left(C_{\alpha a}-M_{\alpha}^{\beta}C_{\beta a}\right),
\end{equation}
with $M_{\alpha}^{\beta}$ a set of $k\times k$ matrix integration
variables. This form makes the integral look more like the Grassmannian
formulation of Ref.~\cite{arkani-hamed_duality_2010}, but more importantly
for our purposes, it allows us to recast it into the form of a contour
integral. The idea is that there are $\left(k-2\right)\left(n-k-2\right)$
delta functions beyond those that fix the kinematics.%
\footnote{In other words, delta functions other than $\prod_{\alpha=1}^{k}\delta^{4|4}\left(C_{\alpha a}\mathcal{W}^{\alpha a}\right)$.%
} These extra delta functions are factors of $G_{\tau}\left(C\right)$.
Following the discussions in Refs.~\cite{spradlin_twistor_2009,arkani-hamed_unification_2011,arkani-hamed_local_2011},
we can reinterpret integrating against one of these delta functions
as instead integrating around a contour that encloses a pole located
at the argument of the delta function. Therefore, using the notation
of Ref.~\cite{arkani-hamed_unification_2011}, the integral can be
recast as

\begin{equation}
A_{n}\left(1,\tau\left(2\right),\dots,\tau\left(n-1\right),n\right)=\int_{S_{1}=\cdots=S_{m}=0}\frac{d^{k\times n}C_{\alpha a}}{\mathrm{vol}\, GL\left(k\right)}\frac{H_{\tau}\left(C\right)}{S_{1}\left(C\right)\cdots S_{m}\left(C\right)}.
\end{equation}
There are $m=\left(k-2\right)\left(n-k-2\right)$ functions $S$,
called Veronese operators, and these contain the locations of the
poles. As the notation suggests, $H_{\tau}$ contains the (integrated)
minor factor $L_{\tau}$, since $L_{\tau}$ was part of $G_{\tau}$.
Therefore $H_{\tau}$ depends on $\tau$, while none of the Veronese
operators do.

A concrete example is the $n=6$, $k=3$ Yang-Mills amplitude. In
this case, there is $m=1$ function $S\left(C\right)$ that determines
the correct contour in the complex plane. Gauge fixing in $GL\left(3\right)$
and overall momentum conservation results in only one remaining complex
integration variable $c$, and so the calculation reduces to a standard
contour integral in $\mathbb{C}$. It may be shown in this $n=6$,
$k=3$ case, that $S\left(c\right)$ is quartic for arbitrary momenta~\cite{spradlin_twistor_2009,arkani-hamed_unification_2011}.
The correct contour for calculating the amplitude must enclose the
four roots $c_{1}$, $c_{2}$, $c_{3}$, and $c_{4}$ of $S\left(c\right)$,
as indicated in Fig.~\ref{fig:Poles}. The remaining three points
$\tilde{c}_{1}$, $\tilde{c}_{2}$, $\tilde{c}_{3}$ in the $c$-plane
correspond to the poles of the function $H\left(c\right)$. 

\begin{center}
\begin{figure}
\includegraphics{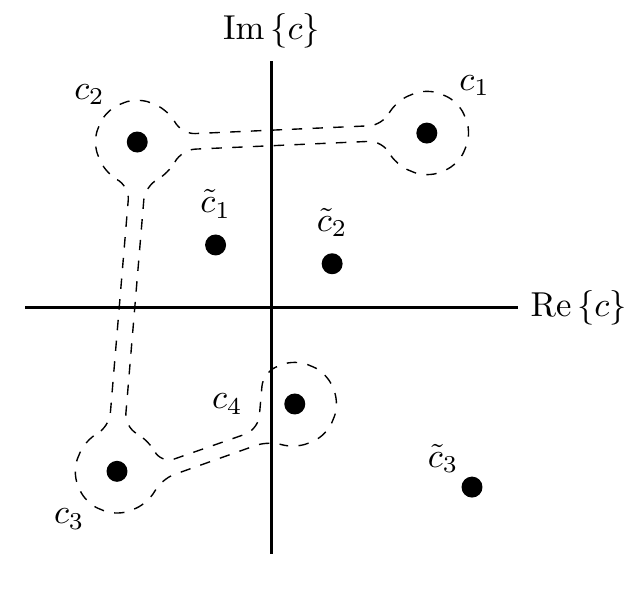}

\caption[Poles of the $n=6$, $k=3$ Yang-Mills amplitude.]{\label{fig:Poles}The integral for the $n=6$, $k=3$ Yang-Mills amplitude
has one integration variable not fixed by gauge choice or momentum
conservation, and so may be calculated as a standard contour integral
of a complex variable $c\in\mathbb{C}$. The four poles $c_{1}$,
$c_{2}$, $c_{3}$, and $c_{4}$ correspond to the four roots of $S\left(c\right)$,
and the three remaining poles $\tilde{c}_{1}$, $\tilde{c}_{2}$,
and $\tilde{c}_{3}$ correspond to the poles of the function $H\left(c\right)$.
This figure is meant only as a guide; the actual location of the poles
changes for different external momenta. }
\end{figure}

\par\end{center}

For general $n$ and $k$, we can then use the global residue theorem%
\footnote{This is necessary because we are dealing with a multidimensional contour
integral. See~\cite{arkani-hamed_duality_2010} for details.%
} to write the amplitude as a sum of residues (where we have absorbed
factors of $2\pi i$):
\begin{eqnarray}
A_{n}\left(1,\tau\left(2\right),\dots,\tau\left(n-1\right),n\right) & = & \sum_{r}R_{r}\left(1,\tau\left(2\right),\dots,\tau\left(n-1\right),n\right).\label{eq:amplitude as residues}
\end{eqnarray}
Letting $C_{r}$ denote the location of residue $R_{r}$%
\footnote{We cannot simply say the $C_{r}$ are the zeros of the Veronese operators,
due to issues at high multiplicity arising from so-called composite
residues. A more complete discussions of these issues can be found
in \cite{arkani-hamed_duality_2010,arkani-hamed_unification_2011}.%
}, the residues have the general form (with the $\tau$-dependence
explicit) 
\begin{equation}
R_{r,\tau}=\frac{H_{\tau}\left(C_{r}\right)}{T\left(C_{r}\right)}.\label{eq:residue definition}
\end{equation}
Here $T\left(C\right)$ is a function consisting of the product of
the nonvanishing Veronese operators and containing the gauge fixing.
Notice that $T$ is independent of $\tau$.

We can now apply the matrix $\mathcal{S}$ to each residue $R_{r}$.
Doing so, we have 
\begin{equation}
\mathcal{S}_{\tau\omega}R_{r,\omega}=\frac{1}{T\left(C_{r}\right)}\mathcal{S}_{\tau\omega}H_{\omega}\left(C_{r}\right).
\end{equation}
Remembering that $H_{\tau}$ is precisely $G_{\tau}$ with some integration
variables fixed by delta functions and the remaining delta functions
stripped off, we see that $H_{\tau}\left(C_{r}\right)$ is proportional
to $L_{\tau}$ evaluated on the support of all of the delta functions.
But, as proved in Ref.~\cite{cachazo_fundamental_2012}, $\mathcal{S}L_{\tau}=0$
on the support of the delta functions in the RSVW formula. Therefore
$\mathcal{S}_{\tau\omega}H_{\omega}\left(C_{r}\right)=0$, and so
$\mathcal{S}$ annihilates the residues $R_{r}$ individually, exactly
as we wished to show.

\subsection{Gravity}

We are ultimately interested in the application of the techniques
developed so far to analyze gravity. We therefore discuss how the
material of the previous two sections can be generalized to gravity.

The first concept is the double-copy formula. This states that tree
level gravitational amplitudes can be written as
\begin{equation}
\mathcal{M}_{n}=\left(\frac{\kappa}{2}\right)^{n-2}\sum_{i=1}^{\left(2n-5\right)!!}\frac{n_{i}\tilde{n}_{i}}{D_{i}},\label{eq:double copy}
\end{equation}
where the $n_{i}$ and $\tilde{n}_{i}$ are both sets of kinematic
numerators, possibly from different Yang-Mills theories (such as with
varying amounts of supersymmetry), at least one set of which is color-dual.
The gravitational coupling constant is $\kappa$. This formula was
first proposed in Ref.~\cite{bern_perturbative_2010}, and was proven
in Ref.~\cite{bern_gravity_2010} using color-kinematic duality and
the BCFW recursion relations. Our ultimate goal is to derive an analogous
formula valid at the level of residues.

The second concept defines exactly what the residues look like on
the gravitational side. The Cachazo-Geyer formula for gravitational
amplitudes was proposed in Ref.~\cite{cachazo_twistor_2012}: 
\begin{equation}
\mathcal{M}_{n}=\int\frac{d^{2n}\sigma}{\mathrm{vol}\, GL\left(2\right)}\frac{H_{n}}{J_{n}}\prod_{\alpha=1}^{k}\delta^{2}\left(C_{\alpha a}\tilde{\lambda}_{a}\right)\delta^{0|8}\left(C_{\alpha a}\tilde{\eta}_{a}\right)\int d^{2}\rho_{\alpha}\prod_{b=1}^{n}\delta^{2}\left(\rho_{\beta}C_{\beta b}-\lambda_{b}\right),\label{eq:Cachazo-Geyer}
\end{equation}
which looks exactly the same as the RSVW formula Eq.~(\ref{eq:RSVW no permutation})
except for the four extra supersymmetries and the replacement of the
inverse minor factor with $\frac{H_{n}}{J_{n}}$. The exact definition
of $\frac{H_{n}}{J_{n}}$ is not important for our purposes, but it
is also a function of the minors of $C_{\alpha a}$. All of the discussion
involving writing the RSVW formula as a contour integral applies to
this formula as well. Indeed, since the delta functions in both formulas
are the same, the residues occur at the exact same points, and the
same contours may be used. This is crucial, as it allows us to put
RSVW residues in one-to-one correspondence with the residues of this
formula. Explicitly, we write 
\begin{equation}
\mathcal{M}_{n}=\sum_{r}R_{r}^{G},
\end{equation}
with the index $r$ matching the RSVW residue index $r$.

The formula Eq.~(\ref{eq:Cachazo-Geyer}) was conjectured in Ref.~\cite{cachazo_twistor_2012};
its proof depended on the resolution of a conjecture called the KLT
orthogonality conjecture. This conjecture was proven in Ref.~\cite{cachazo_scattering_2013-1}.
In addition to putting Eq.~(\ref{eq:Cachazo-Geyer}) on a solid foundation,
it will play an important role in proving the residue analog of Eq.~(\ref{eq:double copy}).
We will give a formal statement of the conjecture at that time.

\section{Residue Numerators\label{sec:Residue-Numerators}}

The lemmas proved in the previous section have important implications.
In particular Eq.~\ref{eq:residue consistency condition}, the consistency
condition for residues, guarantees the existence of what we have christened
\emph{residue numerators}, defined by 
\begin{equation}
N_{r}\equiv F^{+}R_{r}+\left(I-F^{+}F\right)v,\label{eq:residue numerator definition}
\end{equation}
again for arbitrary $v$. These numerators obey all of the properties
of full amplitude numerators, essentially by definition. In particular,
$R_{r}=FN_{r}$.

Recall that the gravity amplitude $\mathcal{M}_{n}$ can be written
in terms of Yang-Mills partial amplitudes using the KLT relations,
which in our notation take the form~\cite{boels_powercounting_2013}
\begin{equation}
\mathcal{M}_{n}=\left(\frac{\kappa}{2}\right)^{n-2}A^{T}F^{+}\tilde{A}.\label{eq:KLT}
\end{equation}
$A$ and $\tilde{A}$ are the partial amplitudes associated with two
(possibly different) Yang-Mills theories. We can now address our central
question: if we combine two $\mathcal{N}=4$ super-Yang-Mills theories
to get $\mathcal{N}=8$ supergravity, can we replace everything by
residues and still get the same result? In other words, we conjecture
\begin{equation}
R_{r}^{G}\overset{?}{=}\left(\frac{\kappa}{2}\right)^{n-2}R_{r}^{T}F^{+}\tilde{R}_{r}.
\end{equation}
To test this hypothesis, we substitute the expressions for $\mathcal{M}_{n}$
and $A_{n}$ as sums of residues into the KLT relations Eq.~(\ref{eq:KLT}).
This yields 
\begin{eqnarray}
\sum_{r}R_{r}^{G} & = & \sum_{r,\tilde{r}}R_{r}^{T}F^{+}\tilde{R}_{\tilde{r}}\nonumber \\
 & = & \sum_{r}R_{r}^{T}F^{+}\tilde{R}_{r}+\sum_{r\neq\tilde{r}}R_{r}^{T}F^{+}\tilde{R}_{\tilde{r}},
\end{eqnarray}
where we have separated out the cross terms in the second line. If
our conjecture is true, it must imply
\begin{equation}
\sum_{r\neq\tilde{r}}R_{r}^{T}F^{+}\tilde{R}_{\tilde{r}}=0.
\end{equation}
This is the aforementioned KLT orthogonality conjecture, and it was
recently proved in Ref.~\cite{cachazo_scattering_2013-1}.

Now we have
\begin{equation}
\sum_{r}R_{r}^{G}=\left(\frac{\kappa}{2}\right)^{n-2}\sum_{r}R_{r}^{T}F^{+}\tilde{R}_{r}.\label{eq:KLT residue sum}
\end{equation}
However, this is insufficient to show the two sides are equal term-by-term.
To show this stronger statement, we need to go back to the derivation
of Eq.~(\ref{eq:Cachazo-Geyer}) found in Ref.~\cite{cachazo_twistor_2012}.
In particular, the residues $R_{r}^{G}$ and $R_{r}$ are dependent
only on the integrands of the RSVW and Cachazo-Geyer formulas. After
inserting two copies of the RSVW formula into the KLT relations, the
use of KLT orthogonality reduces the integral to an integral over
a single set of the RSVW variables (rather than one set for $A$ and
one for $\tilde{A}$). Therefore the integrands, not just the integrals,
are in fact equal. Since the integrands are evaluated at exactly the
same set of points in determining the residues, this implies that
Eq.~(\ref{eq:KLT residue sum}) holds term-by-term, proving $R_{r}^{G}=\left(\kappa/2\right)^{n-2}R_{r}^{T}F^{+}\tilde{R}_{r}$.

We can now recast this in terms of residue numerators. Specifically,
we can write
\begin{eqnarray}
R_{r}^{G} & = & \left(\frac{\kappa}{2}\right)^{n-2}R_{r}^{T}F^{+}\tilde{R}_{r}\nonumber \\
 & = & \left(\frac{\kappa}{2}\right)^{n-2}R_{r}^{T}F^{+}FF^{+}\tilde{R}_{r}\nonumber \\
 & = & \left(\frac{\kappa}{2}\right)^{n-2}\left(\left(F^{+}\right)^{T}R_{r}\right)^{T}FF^{+}\tilde{R}_{r}\nonumber \\
 & = & \left(\frac{\kappa}{2}\right)^{n-2}N_{r}^{T}F\tilde{N}_{r},\label{eq:residue numerator double copy}
\end{eqnarray}
where in going to the second line we have used the fact that the residues
obey the consistency condition $\tilde{R}_{r}=FF^{+}\tilde{R}_{r}$,
and in the last line we have used the fact that for $F$ symmetric,
$\left(F^{+}\right)^{T}$ is also a generalized inverse%
\footnote{Since $\left(F^{+}\right)^{T}$ may be different than $F^{+}$, the
resulting numerators may differ from those generated by $F^{+}$ by
a generalized gauge transformation, but this is irrelevant for out
purposes.%
}. But the last line is precisely the double-copy formula with ordinary
kinematic numerators replaced by residue numerators, and the gravitational
amplitude replaced by the corresponding gravitational residue!

Notice that this argument holds in reverse. Assuming the double-copy
formula for residue numerators, we can reverse the logic in the equations
leading to Eq.~(\ref{eq:residue numerator double copy}) and derive
the KLT relations for RSVW residues. The course of this argument uses
``residue numerator orthogonality''
\begin{equation}
\sum_{\tilde{r}\ne r}N_{r}^{T}F\tilde{N}_{\tilde{r}}=0,
\end{equation}
which follows from the residue numerator double-copy formula, to prove
KLT orthogonality. As a simple consistency check, we have numerically
verified this result at six points ($n=6$ is the smallest $n$ for
which nontrivial residues occur in the connected prescription\cite{spradlin_twistor_2009}).

This equivalency between the KLT relations and the double-copy formula
harks back to the equivalency at the amplitude level~\cite{boels_powercounting_2013},
and we expect it will be important for similar reasons. In particular,
the double-copy formulation has two major advantages over the KLT
relations. First, the double-copy formula is much cleaner, making
the ``gravity is the square of gauge theory'' adage more transparent.
Second, the KLT relations are restricted to tree level, while the
double-copy formula is conjectured to apply to all loop orders. This
suggests that the residue numerators, while only now defined at tree
level, might have loop-level analogs.

We have now shown that not only do the BCJ amplitude identities and
the KLT relations descend from the full amplitude to their residues,
but so do the concepts of kinematic numerators and the double-copy
formula. In the same vein, we emphasize that it is equivalent to use
residue numerators as a starting point, and derive the residue relations,
much as the original BCJ amplitude identities were originally derived
from numerators.

\section{Conclusion and Future Work\label{sec:Conclusion-and-Future-Work}}

We have introduced residue numerators. These objects serve to give
a BCJ decomposition of RSVW residues. They have the property that
under replacement of a color factor by a residue numerator we obtain
gravity residues, and they obey an orthogonality condition equivalent
to the KLT orthogonality condition~\cite{cachazo_scattering_2013-2}.

We have confirmed all three points in a specific case $\left(n=6\right)$.
To generalize this result, we proved two lemmas. First we proved that
the consistency condition of the generalized inverse is equivalent
to the BCJ amplitude identities by counting the rank of the matrices
$F$ and $FF^{+}$, largely in the spirit of the arguments suggested
in Ref\@.~\cite{boels_powercounting_2013}. By replacing amplitudes
with RSVW residues, $A\rightarrow R_{r}$, and $N\rightarrow N_{r}$
in expressions for amplitudes, our proof implies that RSVW residues
obey BCJ amplitude identities if and only if there exist decompositions
of RSVW residues into residue numerators. It is possible to explicitly
construct such residue numerator solutions~\cite{kiermaier_gravity_2010,cachazo_scattering_2013-3},
so we only needed to verify that RSVW residues obey amplitude relations
in general. We did this by demonstrating that RSVW residues contain
permutation-dependent factors which vanish in the BCJ amplitude relation,
as discussed in Ref.~\cite{cachazo_fundamental_2012}. This was the
second lemma that we proved.

From these two lemmas, our main result followed. We showed that the
new proof of KLT orthogonality for RSVW residues implies a residue
numerator orthogonality condition. Conversely, from the ab initio
assumption that both RSVW and gravity residues could be decomposed
into residue numerators, we arrived at a residue numerator orthogonality
condition that implies the KLT orthogonality condition.

We expect that residue numerators will offer insight into a variety
of topics that are currently phrased in terms of amplitudes. At tree-level,
the interplay of combinatorics and linear algebra that go into constructing
$F$ are suggestive of the positive Grassmannian~\cite{arkani-hamed_scattering_2012}
and amplituhedron~\cite{arkani-hamed_amplituhedron_2013-2,arkani-hamed_into_2013}.
The numerator Jacobi relations may be relevant to lifting the theory
described by the amplituhedron out of the planar limit, since numerator
Jacobi identities relate planar and nonplanar diagrams. (See, for
example, Ref\@.~\cite{bern_simplifying_2012}.) Also at tree level,
we suspect a direct link between residue numerators and the recently
constructed scattering equations~\cite{cachazo_scattering_2013-1},
especially in light of the even more recent decomposition of the scattering
equations in terms of kinematic numerators~\cite{cachazo_scattering_2013-3}.
It would also be interesting to explore the role of residue numerators
in theories like ABJM, in which even the tree-level connection between
color-kinematic duality and BCJ amplitude identities is less well
understood~\cite{huang_three-algebra_2013,sivaramakrishnan_color-kinematic_2014}.

Numerator decompositions have proven to be a powerful way to evaluate
loop-level contributions to amplitudes as well. The decomposition
of residues into numerators performed here may offer insight into
improved ways of constructing loop integrands. We expect that reexamining
loop-level amplitudes where known color-dual numerators are available
will offer insight into how residue numerators might be applied in
loop calculations. There are good starting points in the literature
pursuing such systematization at tree level~\cite{fu_algebraic_2013},
at loop level with color-dual numerators~\cite{bjerrum-bohr_integrand_2013},
and at higher-loop level without color-dual numerators~\cite{kosower_maximal_2012,caron-huot_uniqueness_2012}.
Recent work~\cite{bern_color-kinematics_2013} also explicitly illustrates
a color-dual numerator construction for pure Yang-Mills at one loop
and two loops, providing nontrivial evidence that numerator representations
extend to nonsupersymmetric theories. We expect that residue numerators
might be applicable to theories with less supersymmetry.

That said, the extension of the RSVW formula (and more recently the
scattering equations) to loop level has been fraught with difficulties
\cite{adamo_ambitwistor_2013,mason_ambitwistor_2013}. It has been
difficult to understand generalized gauge invariance at loop level~\cite{boels_powercounting_2013},
as the invariance applies at the level of the integrand, and must
therefore be extended to include terms that vanish upon integration
over loop momenta. We hope the algebraic simplicity of residue numerators
will help these barriers be overcome, but at present there is not
an obvious path forward.

Finally, the existence of residue numerators reemphasizes the role
of numerators in color-kinematic duality. In the same way that color-kinematic
duality underlies the BCJ amplitude identities, we have demonstrated
that these residue numerators imply KLT and BCJ relations as well
as the KLT orthogonality relations between RSVW residues.

\section*{Acknowledgments}

We would like to extend our gratitude to our advisor, Zvi Bern, for
suggesting this problem to us and providing invaluable guidance throughout
the course of our research. We would also like to thank Josh Nohle,
Allic Sivaramakrishnan, and especially Rutger Boels, all of whom provided
helpful feedback during conversations. Finally, the referee provided
numerous useful suggestions and improvements, for which we are greatful.
S.L. would like to dedicate his portion of this work to the memories
of his grandfather, Robert Etem, and of his grandmother, Meredith
Litsey. J.S. was supported by the Department of Defense (DoD) through
the National Defense Science \& Engineering Graduate Fellowship (NDSEG)
Program during the majority of this work.

\newpage

\appendix

\section*{Appendix: Obtaining Numerator Residues}

Here we present one possible way of defining numerator residues.

We first argue that six-point next-to-maximal-helicity-violating amplitudes
($\textrm{NMHV}$) present the simplest, non-trivial appearance of
residue numerators. Borrowing from Ref.~\cite{spradlin_twistor_2009},
an amplitude of any helicity may be written as an integral in $\left(k-2\right)\left(n-k-2\right)$
integration variables $c_{i}$: 
\begin{equation}
A_{n}=\int d^{\left(k-2\right)\left(n-k-2\right)}c_{i}\, a_{n}\left(c_{i}\right),\label{eq:amplitude-as-integral}
\end{equation}
where $k$ is the number of negative-helicity gluons in the scattering
process. By definition, an amplitude with $k=2$ is called ``MHV''.
Any amplitude with $k=K+2$ is known as an $\textrm{N}^{K}\textrm{MHV}$
amplitude. The function $a_{n}\left(c_{i}\right)$ in Eq.~\ref{eq:amplitude-as-integral}
is unimportant for the current explanation. These complex integrals
are exactly the ones that produce residues and corresponding residue
numerators. We require that $k>2$ (since $k\le2$ results in no integration
according to Eq.~\ref{eq:amplitude-as-integral}). The simplest case
is then $k=3$. We further look for the case where there is only one
complex integration parameter:
\begin{equation}
\left(k-2\right)\left(n-k-2\right)=1\Rightarrow n=6.
\end{equation}
Thus $n=6$, $k=3$ amplitudes offer the first, simplest opportunity
for the appearance of residue numerators.

While there is likely illuminating structure hiding in the functional
form of the integrands, we here ignore such details in favor of a
broader view. The expression Eq.~\ref{eq:amplitude-as-integral}
for $n=6$, $k=3$ says that any amplitude may be expressed as
\begin{equation}
A_{n}\left(L,h\right)=\int dc\, a_{n}\left(L,h,c\right),
\end{equation}
for a momentum label configuration $L\in\mathcal{P}\left(\left\{ 1,2,\ldots,n\right\} \right)$
and helicity configuration $h=\left\{ h_{1},h_{2},\ldots,h_{n}\right\} $
(the $L$ and $h$ were suppressed in Eq. \ref{eq:amplitude-as-integral}).
For general $n$ and $k$, $a_{n}\left(c_{i}\right)$ contains delta
functions in $c_{i}$. In the case $n=6$, $k=3$, the argument of
the delta function is quartic in the complex variable $c$, and so
the integral may be reexpressed as a contour integral enclosing exactly
the four roots of the argument of the delta function. More explicitly
if
\begin{equation}
a_{n}\left(L,H,c\right)=\mathfrak{a}_{n}\left(L,H,c\right)\delta\left(S\left(c\right)\right),
\end{equation}
then 
\begin{equation}
S\left(c\right)=\kappa\left(c-c_{1}\right)\left(c-c_{2}\right)\left(c-c_{3}\right)\left(c-c_{4}\right)
\end{equation}
for an overall constant $\kappa$. Converting the amplitude into a
complex integral and ignoring factors of $2\pi i$ which cancel out
in the final result:
\begin{align}
A_{n}\left(L,h\right) & =\ointop_{S\left(c\right)}dc\frac{\mathfrak{a}_{n}\left(L,H,c\right)}{S\left(c\right)}\nonumber \\
 & =\sum_{i=1}^{4}\underset{c=c_{i}}{\textrm{Res}}\left(\frac{\mathfrak{a}_{n}\left(L,H,c\right)}{S\left(c\right)}\right)\nonumber \\
 & \equiv\sum_{r=1}^{4}R_{r}.
\end{align}
We now define residue numerators by expressing color-dual numerators
in terms of amplitudes, replacing each amplitude with a residue of
that amplitude, and indexing the resulting residue numerator accordingly.
Schematically:
\begin{equation}
n=f\left(A\right)\implies n_{r}=f\left(R_{r}\right),\ A=\sum_{r}R_{r}.
\end{equation}

There are several methods of determining the function $f\left(A\right)$;
here we present one possible method. We determine residue numerators
in the $n=3$, $k=6$ case by comparing two different expressions
for the gravity amplitudes. The first is the KLT expression at six-point:
\begin{eqnarray}
\mathcal{M}\left(123456\right) & = & -i\left(\frac{\kappa}{2}\right)^{6-2}\sum_{\tau\in S_{3}}s_{1\tau\left(2\right)}s_{\tau\left(4\right)5}\tilde{A}\left(1,\tau\left(2\right),\tau\left(3\right),\tau\left(4\right),5,6\right)\times\nonumber \\
 &  & \hphantom{-i\left(\frac{\kappa}{2}\right)^{6-2}\sum_{\tau\in S_{3}}}\left(s_{\tau\left(3\right)5}A\left(\tau\left(2\right),1,5,\tau\left(3\right),\tau\left(4\right),6\right)+\right.\nonumber \\
 &  & \hphantom{-i\left(\frac{\kappa}{2}\right)^{6-2}\sum_{\tau\in S_{3}}}\left.+\left(s_{\tau\left(3\right)\tau\left(4\right)}+s_{\tau\left(3\right)5}\right)A\left(\tau\left(2\right),1,5,\tau\left(4\right),\tau\left(3\right),6\right)\right),\label{eq:gravity KLT}
\end{eqnarray}
where $S_{3}$ is the set of all permutations of $\left\{ 2,3,4\right\} $.
The second is the numerator decomposition of the gravity amplitude
\cite{bern_gravity_2010}:
\begin{equation}
\mathcal{M}\left(123456\right)=i\kappa^{6-2}\sum_{\tau\in S_{4}}n_{1\tau\left(2\right)\tau\left(3\right)\tau\left(4\right)\tau\left(5\right)6}\tilde{A}\left(1,\tau\left(2\right),\tau\left(3\right),\tau\left(4\right),\tau\left(5\right),6\right),\label{eq:gravity numerator decomposition}
\end{equation}
where $S_{4}$ is the set of all permutations of $\left\{ 2,3,4,5\right\} $.
Equating the two expressions for $\mathcal{M}\left(123456\right)$
given in Eq.~\ref{eq:gravity KLT} and Eq.~\ref{eq:gravity numerator decomposition}
yields expressions for the $\left(n-2\right)!$ numerators $n_{1\tau\left(2\right)\tau\left(3\right)\tau\left(4\right)\tau\left(5\right)6}$:
\begin{eqnarray}
n_{1\tau\left(2\right)\tau\left(3\right)\tau\left(4\right)56} & = & -2^{-4}s_{1\tau\left(2\right)}s_{\tau\left(4\right)5}\left(s_{\tau\left(3\right)5}A\left(\tau\left(2\right),1,5,\tau\left(3\right),\tau\left(4\right),6\right)\right.\nonumber \\
 &  & \hphantom{-2^{-4}}\left.+\left(s_{\tau\left(3\right)\tau\left(4\right)}+s_{\tau\left(3\right)5}\right)A\left(\tau\left(2\right),1,5,\tau\left(4\right),\tau\left(3\right),6\right)\right),
\end{eqnarray}
\begin{equation}
n_{1\tau\left(2\right)\tau\left(3\right)\tau\left(4\right)\tau\left(5\right)6}=0\ \left(\textrm{for }\tau\left(5\right)\ne5\right).
\end{equation}
The residue numerators are then constructed by replacing 
\begin{equation}
A\left(\tau\left(1\right),\tau\left(2\right),\tau\left(3\right),\tau\left(4\right),\tau\left(5\right),\tau\left(6\right)\right)\rightarrow R_{r}\left(\tau\left(1\right),\tau\left(2\right),\tau\left(3\right),\tau\left(4\right),\tau\left(5\right),\tau\left(6\right)\right),
\end{equation}
where Eq.~\ref{eq:amplitude as residues} holds. Explicitly:
\begin{eqnarray}
n_{r,1\tau\left(2\right)\tau\left(3\right)\tau\left(4\right)56} & = & -2^{-4}s_{1\tau\left(2\right)}s_{\tau\left(4\right)5}\left(s_{\tau\left(3\right)5}R_{r}\left(\tau\left(2\right),1,5,\tau\left(3\right),\tau\left(4\right),6\right)\right.\nonumber \\
 &  & \hphantom{-2^{-4}}\left.+\left(s_{\tau\left(3\right)\tau\left(4\right)}+s_{\tau\left(3\right)5}\right)R_{r}\left(\tau\left(2\right),1,5,\tau\left(4\right),\tau\left(3\right),6\right)\right).
\end{eqnarray}

This approach may seem circular since the residues are used to define
the residue numerators. In the end, however, the residue numerators
are nothing more than complex numbers $n_{r,1\tau\left(2\right)\tau\left(3\right)\tau\left(4\right)\tau\left(5\right)6}\in\mathbb{C}$
that serve as the numerators for the residues of amplitudes, and the
manner of determining those complex numbers is irrelevant.

\newpage

\bibliography{Residue_Numerators_Paper_-_August_2013}

\end{document}